# From Ontology to Structured Applied Epistemology

Robert B. Allen
Yonsei University, Seoul, South Korea
`rballen@yonsei.ac.kr`

**Abstract:** Developing and organizing new knowledge is a core activity for scholars. Recently, ontologies have been introduced as an approach for organizing knowledge. However, most ontologies do not readily support the development and organization of new knowledge. By comparison, to ontology, epistemology is the study of what can be known. Aspects of epistemology include the acquisition of and justification for new knowledge. Thus, we need to coordinate ontology with epistemology. Because we are developing frameworks for capturing knowledge across several scholarly domains, we describe the work in this paper as exploring structured applied epistemology. Unlike other recent proposals for new approaches to scholarly publishing, we propose an integrated and comprehensive approach. We have explored "direct representation" based on the rigorous Basic Formal Ontology and in this paper, we consider how epistemology can be incorporated with that. In addition to highly-structured scientific research reports, we also consider how to develop highly-structured descriptions of historical events on which historical analyses can be based.

## 1 Introduction

We have been exploring representations for highly structured research reports (e.g., [2, 3, 4, 5]) and community models (e.g., [9, 23]). In addition to modeling specific activities, we have also considered broader issues on "model-oriented information organization" (e.g., [6, 7]). While ontologies are most often used to provide descriptive labels, we emphasized using them for "direct representation" by which we mean using them for developing detailed semantic models.

Because, such models need to be based on rigorous semantics, we have adopted the Basic Formal Ontology (BFO) that is widely used in biology and is being considered as an ISO standard. However, scholarly work goes beyond description of the Entities in a domain. Research reports often make claims about how those domain ontologies should be changed. Here, we consider several steps for structuring the description of those claims, for evaluating them, and ultimately for merging them into the broader knowledge base. Effectively, these activities aim at developing and justifying new knowledge. Rather than the usual application of an ontology to support reasoning about a domain, our work asks how can we determine what is the best ontological structure that can be applied to the domain. We call this type of project "structured applied epistemology". Because, structured revisions to knowledge are integral to scientific research and analytical histories, such reports and descriptions should support the ability to describe, justify, and ultimately to implement changes to knowledge models.

The assessment and incorporation of new knowledge implies that the ontology must change, which can be considered a type of learning. Figure 1 adapts the well-known model of double-loop learning [14, 26] to the use and revision of semantic models. The output of the semantic model is compared to what is known about the real world. A mismatch may trigger an effort to identify the source of the discrepancy and then to update ontologies.

Several different strategies may be applied to address that mismatch. The details of the Instance Model may be changed. These many include identifying or setting values of specific attributes or changing the assignment of entity types. Or, the transitions which cause those changes may be updates. In other cases, the mismatch may suggest that the domain ontology needed to be revised or extended. Very rarely, there might also be changes to the upper ontology. The strategies for determining whether revisions are needed is applied epistemology.

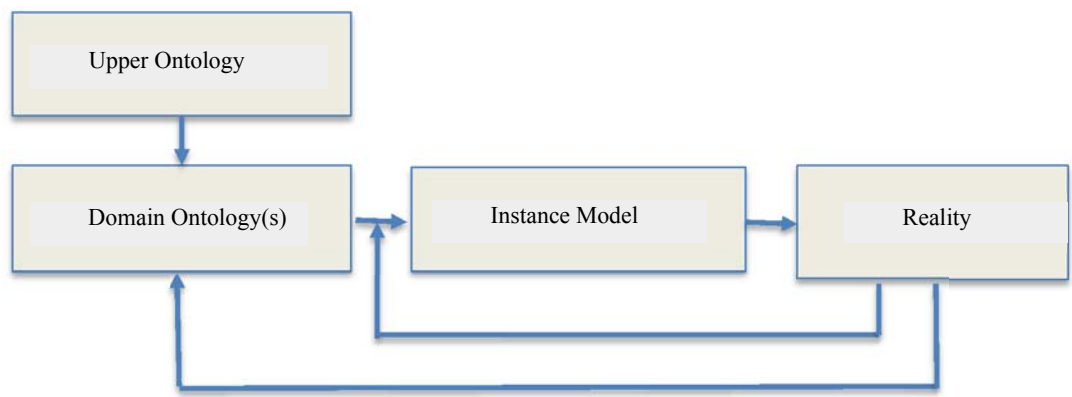

**Figure 1: Double loop ontology learning**

Section 2 examines ontologies and Section 3 examines epistemology. Section 4 considers structured applied epistemology for scientific research reports while Section 5 considers structured applied epistemology for history. Section 6 explores next steps.

## 2 Ontologies

Ontologies can provide structure and coherence to ad hoc linked data. They propose entities and relationships among those entities that provide a description of a domain and they generally support hierarchical *is_a* relationships among the entities. Our recent work has focused on applying the Basic Formal Ontology (BFO2.0) [15]. An upper ontology provides semantic data typing for domain ontologies derived from it.

BFO claims to be a "realist ontology which means it describes only entities that exist. Realist ontologies encourage what we have called "direct representation". That is, its entities describe in reality directly rather than describing them indirectly. The main components of BFO are Continuants (3D) and Occurrents (4D). The Continuants are divided into Independent Continuants and Dependent Continuants. Dependent Continuants are further divided into entity types that include Qualities and Realizables. Entities are connected by relationship and there are several different types of relationships. Rather being a single comprehensive ontology, the BFO is better thought of as a family of interrelated ontologies [15, chp 2]. Thus, the application of the different ontologies in different situations involves a type of scoping (cf., [32, 40]). The two levels in Figure 1 correspond roughly to the distinction between Universals and Particulars.

### 2.1 Extending BFO with Rich Structures and Rich Semantics

BFO provides a solid foundation for our application. However, as we explore how it can be applied to highly-structured information organization we see that there are several areas where extensions are needed. While BFO, and most other ontologies, are based on sets of triples, higher-level structures are also needed.

As noted above, there are several different types of entities in BFO. For instance, Material Entities (a type of Independent Continuant) are distinguished from their Qualities (e.g., Color, Mass). Especially when we are talking about things in the world, it is helpful to have a term that describes the bundle of an Independent Continuant and it Dependent Continuants. This bundle has been called a template or composite; it could also be seen as a "frame".

As part of the Referent Tracking initiative, [20] proposes assigning an IUI to entities and their attributes (i.e., Dependent Continuants) such as individual patients as described by EHRs. In later work, the authors also propose applying IUIs to information resources described with DOIs and things in Internet of Things.

Allen [9, 10, 12] has also examined composite entities and suggests including their include their sub parts. Consider, for instance, applying semantic modeling as an approach to whole-cell or whole-organism modeling [34]. Such composites are often systems and can be said to have "causal unity". Different types of systems may be identified; some systems simply maintain an equilibrium while others have a function in the context of broader systems.

In addition, we define a Scenario as a composite entity or group of entities which interact across time.

### 2.2   States and State Transitions

While BFO supports direct representation of static domains, it does not directly support dynamic domain descriptions. In particular, it does not currently include the notion of State although the motion of Composite Entities described above, allows the specification of it State [17] (cf., [9, 12]). We can represent State Changes as Transitions. In other words, a State Change occurs every time a relationship changes. Processes in BFO are Occurrents in BFO which "occur" or "happen" in time [15, p183]. However, it is not clear how Processes are related to Transitions.

Ultimately, we believe that Transitions should be directly related to type of State Change they cause. For instance, some Transitions cause changes in spatial location while others cause changes in Qualities such as color or in Roles. Higher level Processes would be composed of these action primitives which are directly related to the structure of the BFO ontology. It is interesting to note that many prepositions in natural language are related to these structures. There are prepositions for spatial location (e.g., "in the house") and for thematic relatedness [aboutness] (e.g., "about the author"). In addition, many natural language verbs (e.g., to own, to eat) imply the existence of specific properties and transitions [8, 43].

### 2.3   Descriptive Programs

It seems surprising that that there has not been a greater effort to integrate ontologies with programming languages. There are many reasons this would be useful. Programing languages have features such as looping and conditionals that provide representational power beyond what is possible with typical ontologies. Moreover, higher-level program structures such as workflows and procedures (e.g., [13, 22]) can be easily implemented.

More broadly, programming is closely related to the notion of a system (Section 2.1). If we consider the description of a Scenario as a type of systems analysis, then we can see the relevance of software development methodologies. This task is also related to object-oriented modeling but the goal is not the design of the system, rather it is the development of a descriptive framework.

The composite entities described in Section 2.1 can be related to classes in a programming language [8, 9, 12, 17, 22]. Given an object-oriented approach, the Processes associated with the composite classes can be modeled as methods. In addition, the interaction of separate classes can be viewed as similar to interacting agents in a Scenario.

Such an approach is compatible with "model-oriented information organization" [6, 7]. It also supports the notion of "program-based ontologies" [17] applies the functional programming language Hasekell for validation of updates made during Referent Tracking (see Section 2.1; also see [8, 12]).

## 3 Epistemology

In the broadest sense, epistemology is concerned with what can be known. Somewhat more narrowly, epistemology is concerned with how knowledge (or "true belief") is acquired and justified. We are interested in formal systems and structures for epistemology because we aim to develop information systems that combine semantic models with evidence, warrants, argumentation [45], and claims regarding the acquisition of knowledge. An interactive system should support browsing the details of evidence, warrants, and claims. For claims based on scientific experiments, we should be able to browse a structured model of the experiment(s) that support those claims (see Section 4). Similarly, we should be able to browse a structured model of entities and events for social histories.

The term "discourse" is related to the use of language for interaction between people. Establishing true beliefs about the world is central to human communication. These are promising steps that we believe can now be integrated into a comprehensive system.

### 3.1 Discourse Elements

Linking of epistemological evidence with specific processes demonstrates the importance of insisting that the semantics and the discourse are directly coordinated. Just as we emphasize direct representation of knowledge, we propose that discourse should be direct. Rather than simply saying that a textual proposition "has a claim", a structured ontological representation should "make a claim" or "make an argument".

Evidence should be on a reliable method for producing true beliefs. Such "reliabilism" is often associated with establishing causal processes [31]. For scientific research, presumably when well executed, Randomized Controlled Trials are reliable method but others are less clear. This issue is addressed in the discussion of research designs and their validity (Section 4.2).

### 3.2 Discourse Vocabularies

Several discourse vocabularies have been proposed and in some cases, these have been extended as ontologies (see Section 3.2). The discourse vocabularies apply to the various components of discourse claims, argumentation, and evidence. As we shall see in the following section, several of these have been incorporated into ontologies or otherwise applied to information management.

One well-known discourse vocabulary is from Rhetorical Structure Theory [38]. Its terminology includes not only "evidence" and "justification" but it also includes a rich set of terms that would be useful for argumentation such as "concession", "condition", and "antithesis". Although it is not yet well developed for supporting structured argumentation, RST would seem to have considerable potential for that. However, RST does not support the coordination of discourse tags with rich semantics. For instance, RST defines evidence as "something that is intended to increase a reader's belief" [38, p251].

Toulmin's approach [50] does not have as rich a vocabulary as RST. It focuses only on positive examples; but it is notable in proposing a rich structure that describes a specific relationship among the different elements (cf., [41]).

### 3.3 Using Vocabularies to Support Structured Scholarly Discourse

Several discourse vocabularies have been incorporated into information systems that support scholarship. One of the earliest such frameworks was Issue-Based Information System (IBIS) [51]. The graphical hypertext interface, gIBIS [25], was developed to support that system. gIBIS led to the implementation of a number of other composite hypertext argumentation systems. Such hypertext structures are said to have "typed links" where the link types can be different discourse relationships.

However, hypertext oriented work has declined and recently been replaced by vocabularies which are incorporated into ontologies. ScholOnto [46] is an interesting early scholarly discourse ontology but it does not apply our principle that the ontology should be coordinated with the semantics of the article. The SWAN ontology [23] is notable but the links it provides are mostly at the level of entire research reports, or perhaps, for sections of research reports.

The Argumentation Interchange Framework (AIF) [21, 42] provides structures related to argumentation dialogs, a sort of argumentation grammar. AIF has been proposed as an argumentation framework by [18]. Some elements of AIF are quite promising such as allowing that the quality of evidence can be affected by determining whether its source is reliable. However, detailed support is needed for those judgements and it should be tied back into the "unified temporal map" of instances [13]. In short it does not seem likely that AIF is truly compatible with the spirit of micro-publications.

While we do not propose a specific discourse vocabulary here, we can reiterate our principles of direct representation and the inclusion of specific justifications. Thus, we might say that a justification was inconsistent with the relevant domain models. Indeed, such detailed assertions are common in scientific research reports. Moreover, many of these discourse ontologies are limited to sets of triples without supporting a higher-level framework.

# 4 Toward Highly Structured Scientific Research Reports

Science is a systematic approach to developing new knowledge and scientific communication provides a type of structured applied epistemology. It is now widely accepted that scholarly publications are changing. Much of this change is due to new opportunities with digital production and dissemination but an important part of the change is driven by the potential of so-called semantic publishing [43] in which terms from ontologies are used as semantic annotations. Scientific literature has become increasingly structured across time. We take this trend to the extreme and propose that research reports should be totally structured.

Because the writing in scientific research reports is very specific, we propose that essentially every statement in a research report has an identifiable purpose and its content contributes to the overall value of the research. It is now possible to envision research reports in which essentially every statement is structured.

## 4.1 Frameworks for Scientific Communication

The clearest description of traditional structure for research reports was by Swales [48] specification of several aspects of science who described it as IMRD. It is apparent that IMRD is more an ideal than an absolute standard. Over the years, several proposals have been made for loosening but not eliminating it structure. Particularly notable was ([27], sec 5). This is consistent with our proposals supporting navigation within highly structured research reports [5]. One notable proposal for that work was the suggestion for de-emphasizing traditional citations and, instead, directly linking of related constructs.

"Nano-publications" [33] are recent proposal is for very short descriptions of research findings which include just a few attributes. What's more, several proposals have been made for adding links about evidence [24] and argumentation [18] and creating "micro-publications". However, we believe that unified rich structures incorporating models, claims, evidence, warrants, and instance data will eventually be needed. Thus, we believe that "micro-publications" will eventually must be embedded in the broad context of a highly structured knowledgebase.

## 4.2 Structured Components of Model-Based Research Reports

Following IMRD, each section of the research reports have a distinct purpose and those structures can be captured.

**Introduction:** The main activities in the Introduction is describing the Research Motivation, the Research Question, and identifying the gap in the previous literature. Or, as Swales describes it Creating a Research Space (CARS). The research space seems best characterized not as a gap but as an assembly of relevant findings and model elements that may be useful to the effort. We also note that unlike most models, we must allow explicit placeholders for unknowns.

**Method:** While the method section should include descriptions of procedures implementing any manipulation and for recording the results, we believe that the focus of the method section is the specification of the research design (e.g., [19]).

**Results:** The results of the research include not only the main dependent variables but also a description of the execution of the research. However, again, we focus on the Research Design; A research design is more than a simple workflow its relationship to the goals and implementation of the research must be validated. Typically, internal validity is distinguished from external validity [19]. Internal validity is associated with whether the research implementation accomplished what it set to do. For instance, internal validity might examine checks on the manipulations. External validity is, typically, based on whether the research addresses modeling is relevant to the research gap as described in the Introduction. In terms of the research epistemology, we believe that a validated research design can be considered as a warrant.

**Discussion:** The discussion reviews the claims of the research and may propose revisions to the ontologies based on those claims. Further, it may highlight the implications of any revisions that may be indicated. However, we believe that formal changes in standard ontologies will require community consensus.

### 4.3 Reality Check against Contemporary Research Reports

We believe that it is helpful to compare our proposals for descriptive frameworks with actual research reports. While IMRD is widely recognized, even quick consideration suggests that it is often not followed. For instance, journals in the Public Library of Science (PLoS) include many of the research procedures as an appendix to the main article. We picked a specific article [52] for detailed examination (also [5]). We also found that much of the article dealt with the construction and validation or research techniques rather than with directly testing claims. Rather than following a simple IMRD structure, [52] had many sub-studies conducted which often provided convergent evidence. Moreover, we found that identifying and manipulating potential causal relationships were central (cf., [29]). Overall, these observations highlight the need to coordinate the semantic models closely with the discourse relationships.

## 5 Toward Highly Structured Descriptions of Social History

In a broad sense, histories many include any structured description of entities and events. This may include scholarly materials [13] and Referent Tracking for EHRs [20]. However, in this paper we focus on social history such as with community models [8], societal models, narrative histories, and analytical histories [11].

While history is generally focused on instances and there are few, if any, "covering theories", historical modeling still involves a range of types of modeling. For instance, in some histories rich constructs such as trends, generalizations are common. In addition, evidence and justification are needed as much for history as for science although the criteria are different. Inference about social networks and the characteristics of social groups based on fragments of evidence (i.e., prosopography) are particularly common.

Until recently, most "event ontologies" such as ABC Harmony [36] and CIDOC [28] have focused on the events association with the descriptions of entities rather than with changes in the entities themselves. While that is changing (e.g., [1, 30]), event ontologies still are not based on upper ontologies.

Some other recent work on modeling history attempt to describe historical people and events directly but that work is also, largely, based on ad hoc ontologies [39]. Perhaps our proposals (Section 2) will allow the development of event ontologies based on the BFO. Because we are generally focused on describing events as they have occurred in the world, the application of ontologies to them may be seen as reverse engineering of what models world have produced such results.

There will be many advantages to structured descriptions of history based on a consistent upper ontology along with structured applied epistemology. For instance, recently, there has been work on Document Acts and Social Acts that are compatible with the BFO [22].

In addition, we could develop a project that would be parallel to the Open Biomedical OBO Foundry project [48] and leverage some of its tools, but would support the development of ontologies needed for community models.

# 6 Next Steps and Conclusion

Establishing causation for history is fraught. The standards of evidence in history [35] are related to those in law and journalism. Any interpretation will likely be open to dispute. Ultimately, probably the best that can be done is to attempt to apply Simmel's "ideal line" (just as we might apply Occam's Razor for science). Thus we need to develop vocabularies and tools for supporting that. Similarly, we should support participatory argumentation.

We have considered several aspects of coordinating the BFO upper ontology with epistemology as applied to highly structured scholarly research reports and to social histories. We also note that people use composite entities and epistemology in everyday descriptions of the world. The approach presented here may be extended more broadly to modeling natural language.

We envision that the structured research descriptions would be managed as a digital library. In addition, the rich the research reports and historical analyses may result in updates to the ontologies. However, such updates would not be done lightly and would often be based on a community consensus (e.g., [37]). Indeed, it is quite possible that there would be multiple master ontologies and several those might maintain updates at different levels of confidence. The records of the community discussion as to whether be made in the master ontologies needed to be changes should be recorded and preserved. Furthermore, beyond support for the management of the structured descriptions and ontologies, end user services, such as interactive tutorials, may also be able to be developed.